\documentstyle[11pt,paspconf,epsf]{article}

\begin{document}

\title{Implications of H$\alpha$~Observations for Studies of the Cosmic Microwave Background}

\author{P. R. McCullough\altaffilmark{1}}
\affil{Astronomy Department, University of Illinois at Urbana-Champaign,
    Urbana, IL 61801}

\author{J. E. Gaustad}
\affil{Physics \& Astronomy Dept., Swarthmore College, Swarthmore, PA, 19081-1397}

\author{W. Rosing}
\affil{Las Cumbres Observatory, Los Gatos, CA 95033}

\author{D. Van Buren}
\affil{IPAC/Caltech, Pasadena, CA 91125}

\altaffiltext{1}{Cottrell Scholar}

\begin{abstract}
We summarize the relationship between the free-free emission foreground
and Galactic H$\alpha$~emission. 
H$\alpha$ observations covering nearly the entire celestial sphere are
described. These data
provide a template to isolate and/or remove the effect of Galactic
free-free emission from observations of the cosmic microwave
background. Spectroscopy and imaging provide two complementary methods
for measuring the H$\alpha$~emission. Spectroscopy is favored for its
velocity resolution and imaging is favored for its angular resolution.
With templates of the three dominant foregrounds, namely
synchrotron, thermal dust, and free-free emissions, it may be possible
to quantify the location and the brightness of another foreground
such as that from rotating dust grains.

\end{abstract}



\section{Introduction}

Cosmologists interested in the anisotropy of the Cosmic Microwave
Background (CMB) must observe from
within the Milky Way.  Thus, they must measure and subtract foreground
emissions from the Milky Way.
Three important components of foreground
emission have long been recognized: 1) thermal emission from dust, 2)
synchrotron emission from relativistic electrons accelerated by
magnetic fields, and 3) bremsstrahlung (``braking'' radiation or
``free-free'' emission) from
free electrons accelerated by interactions with ions in a plasma. 
A fourth component associated with rotating dust has also been proposed
recently (Draine \& Lazarian 1998).

This review addresses the Galactic free-free emission and its
tracer H$\alpha$~light.
It treats the warm ionized medium (WIM) as an unavoidable foreground
that must be accounted for in the interpretation of microwave observations of
the CMB. 
That is, it treats the WIM phenomenologically
and does not discuss why the WIM is the way it is.

From that viewpoint, the problem with free-free radiation is two-fold.
First, whereas at low
frequencies ($\nu \la 20$~GHz) synchrotron emission dominates, and at
high frequencies ($\nu \ga 100$~GHz) dust emission dominates, in
between ($\nu \approx 30$~GHz)
free-free emission is as bright or brighter than the other two.
That is, we have leverage on synchrotron and dust, but not
free-free.
We need a tracer of free-free to solve the first problem (lack of leverage),
and that tracer is H$\alpha$~light.
The second problem has been -- until very recently --  the lack of
all-sky observations of H$\alpha$. The H$\alpha$~data that solve the second problem
now exist and are being calibrated and rendered into
a template suitable for decontaminating the free-free foreground.

This review is organized as follows. The next section is
a short historical vignette to give context to modern efforts.
Section 3 lists other reviews similar to this one. 
Section 4 relates H$\alpha$ and free-free emissions in Equation 4,
Figure 1, and Table 1.
Section 5 describes and compares
the two principle techniques for observing H$\alpha$, Fabry-Perot spectroscopy
and narrowband imaging. Typical results are shown and inherent
challenges are noted. Section 6 addresses the
association of dust and ionized gas. Finally, Section 7
concludes with a short summary of how the H$\alpha$~data will
be used in practice to decontaminate the microwave observations.

\begin{figure}
\plotone{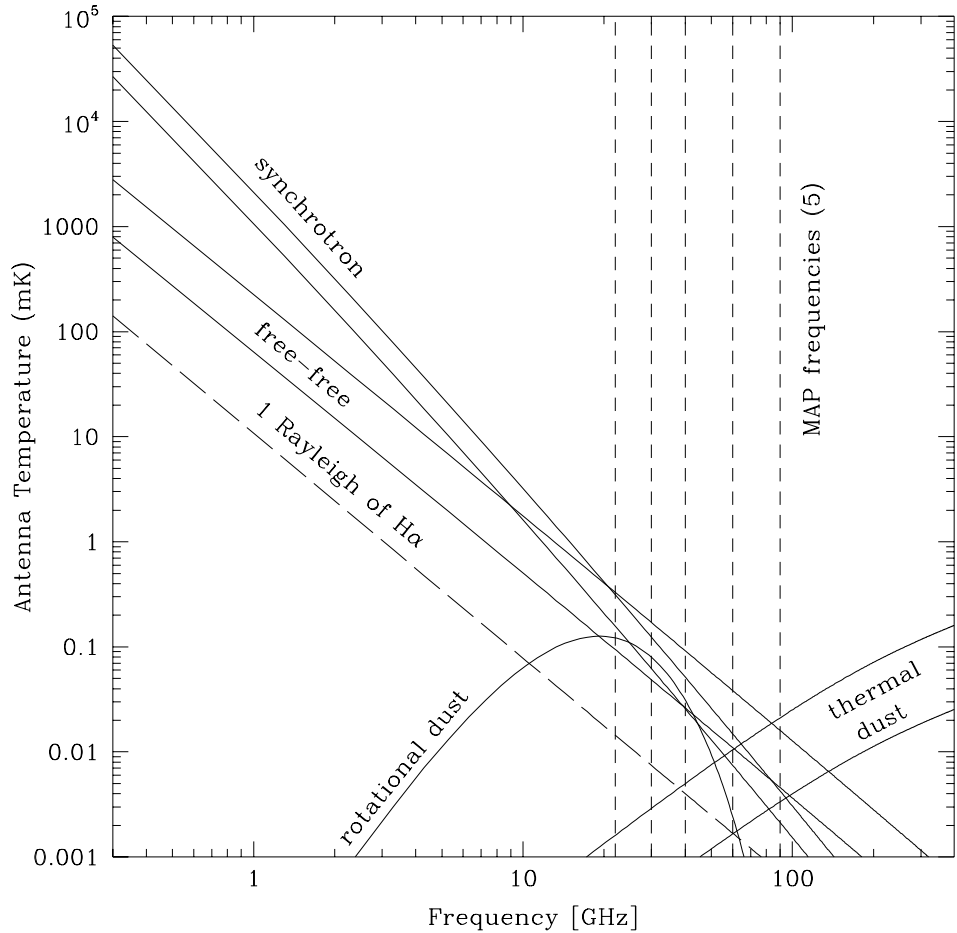}
\caption{ \label{mechanisms}
Foreground contaminants of the CMB are
plotted along with the five bands of the MAP satellite.  At low
frequencies ($\nu \la 20$~GHz) synchrotron emission dominates, while at
high frequencies ($\nu \ga 100$~GHz) dust emission dominates,
and in between free-free emission is as bright or brighter than the other two.
For the synchrotron, thermal dust, and free-free bands,
the upper line corresponds to $|b|\approx 15$\deg, and
the lower line corresponds to $|b|\approx 65$\deg.
The bands are reproduced from Bennett et al (1992).
The Bennett {\it et al.} model has more free-free emission than
models based upon H$\alpha$ observations (cf. Equations 2 and 4).
The sloped, dashed line illustrates the free-free emission corresponding
to an H$\alpha$ surface brightness of 1 Rayleigh.
The curved line illustrates the emission from rotating
dust predicted by Draine \& Lazarian (1998) for the North
Celestial Pole region studied by Leitch {\it et al.} (1997).
The five frequencies of the MAP satellite are 22, 30, 40, 60, and 90 GHz.
} \end{figure}

\section{Yesteryear} \label{yesteryear}

It is interesting to compare today's remote-- and robotic--observing 
in Section 5 with heroic efforts from a century ago.
E. E. Barnard claims to have seen {\it with his own eye} that
which we now call ``interstellar cirrus.''
He photographed filaments of dust in the Taurus region, 
far from the bright reflection nebulosity of the Pleiades,
and remarked
``it was the knowledge of their existence by visual observations alone
that led me to make the photographs'' (Barnard 1896).
For one such photograph, a 10-hour exposure, Barnard kept the plate
in the camera for over fifty hours. 
Unsatisfied with the duration of the exposure at the end of the first night,
he waited for a second night, but it was cloudy, so he resumed the
exposure the third night, after re-aligning the guide star
to its original location.
Barnard's photographs of the filaments streaming away from the Pleiades
look astonishingly like
the 100~\micron~image of IRAS (cf. plate 10 of Barnard (1900)
or plate 8 of Barnard (1928), which is in Arny (1989), to the IRAS
image of Souras (1989)).
Those same filaments absorb background H$\alpha$ (Figure 5), which is
unusual at moderate Galactic latitudes and contrary to the positive correlation
between H$\alpha$ and dust discussed in Section 6.

The H$\alpha$  line is included in the pass band for the red plates
of the National Geographic Society/Palomar Observatory Sky Survey (POSS
1), and these show very bright emission nebulae. Because they
lack the sensitivity to detect the low-surface-brightness WIM,
the POSS plates can give a false impression that ionized gas exists only in
Stromgren spheres, when in fact most (90\%) of the Milky Way's ionized gas
is outside individual Stromgren spheres around OB stars (Reynolds 1993).
The next generation plates (POSS 2) are sensitive to emission of $\sim 25$~R,
or $\sim$4 times fainter than POSS 1, but for both POSS 1 and POSS 2, it
is impossible to identify and quantify the emission specifically from
the H$\alpha$  line because the filters are broadband.

Sivan (1974) conducted a photographic survey in search of
the WIM, reaching a limiting brightness of 15 R and
covering low Galactic latitudes ($|b| \la$ 20\deg) at a resolution
of approximately 10\arcmin.  Later, Parker, Gull, and Kirschner (1979),
using an image tube camera, produced an atlas 
of images in four emission lines ([SII],H$\alpha$  +[NII],
O[III], H$\beta$) and the blue continuum (4220 \AA), which covered low
Galactic latitudes ($|b| \la 8$\deg) at an angular resolution of
0.6\arcmin~and sensitivity of 15 R.

\section{Other Reviews} \label{others}

In this section I note briefly recent reviews similar to
this one. Each of the other reviews describe the physical
relationship between the H$\alpha$  and free-free emissions
and their implications for observing the CMB.
Smoot (1998) reviews the topic as it was in January, 1998.
Valls-Gabaud (1998), Bartlett \& Amram (1998), and Marcelin {\it et al.} (1998)
are similar to each other, each with a slightly different emphasis.
(Valls-Gabaud, Bartlett, and Amram are co-authors of Marcelin {\it et al.}).
Valls-Gabaud (1998) is notable for its thorough discussion of the
proportionality between the brightness temperature due to free-free
$T_b^{ff}$ and the surface brightness in H$\alpha$, $I_\alpha$.

This review necessarily repeats some of the material in the aforementioned
reviews and adds to them the advances of 1998, by the end of which
the H$\alpha$ sky had been observed very nearly in its entirety.

\section{The free-free/H$\alpha$  ratio} \label{recomb}

Because there are treatises on the topic of electron-ion
interaction (e.g. Mezger \& Henderson 1967; Osterbrock 1989, Jackson 1975),
and because the papers
cited in the previous section summarize the topic, we do not
repeat that information here except to note the most basic formulae.

Both free-free and recombination radiation (e.g. H$\alpha$~)
are proportional to the emission measure, 
\begin{equation}
EM = \int n_e n_p dl \approx \int n_e^2 dl.
\end{equation}
At moderate Galactic latitudes, the H$\alpha$  emission (which traces the free-free emission)
has been modeled as (Reynolds 1992)
\begin{equation}
I_\alpha \approx 1.2 R / sin|b|
\end{equation}
where 1 Rayleigh (R) is defined as
\begin{equation}
{\rm 1~R = 10^6/4\pi~photons~cm^{-2}~s^{-1}~sr^{-1},}
\end{equation}
which for H$\alpha$  corresponds to
an emission measure EM = 2.75 cm$^{-6}$pc for gas at 10$^4$~K (Reynolds 1990),
or $2.4\times10^{-7}$ erg cm$^{-2}$ s$^{-1}$ sr$^{-1}$,
or
$5.7\times10^{-18}$ erg cm$^{-2}$ s$^{-1}$ arcsec$^{-2}$.
Like the other constituents of the Milky Way, the H$\alpha$~
varies considerably across the sky.

The ratio of free-free brightness temperature $T_b^{ff}$ to the
H$\alpha$  surface brightness $I_\alpha$ for Case B recombination
is (Valls-Gabaud 1998) 
\begin{equation}
{{T_b^{ff} [mK]}\over{I_\alpha [R]}} = 10.4 \nu^{-2.14} T_4^{0.527} 10^{0.029/T_4} (1 + 0.08)
\end{equation}
where $\nu$ is the observing frequency in GHz, and
the last factor (1+0.08) assumes that helium
(0.08 by number with respect to hydrogen)
is entirely in the form of He$^+$, which creates free-free emission
like hydrogen, but of course doesn't emit H$\alpha$  light.
This relation is the basis for using H$\alpha$  as a tracer of free-free.
It plotted in Figure 1 and is tabulated below.
 
\begin{table}
\caption{Free-free to H$\alpha$  ratio} \label{tbl-1}
\begin{center}
\begin{tabular}{lcc}
Satellite & Frequency [GHz]& ${{T_b^{ff}}/{I_\alpha}} [~{\mu}K~R^{-1}~]$\\
\tableline
COBE	& 19.2	& 19.3	\\
    	& 31.5	& 6.7	\\
    	& 53  	& 2.20	\\
    	& 90  	& 0.71	\\
MAP 	& 22  	& 14.4	\\
    	& 30  	& 7.4	\\
    	& 40  	& 4.0	\\
    	& 60  	& 1.69	\\
    	& 90  	& 0.71	\\
\end{tabular}
\end{center}
\tablenotetext{~}{Values are for $T_e = 8000$~K. Multiply them by 0.88, 0.94, 1.05, or 1.11 for $T_e =$ 6000, 7000, 9000, or 10000 K, respectively. Case B recombination is assumed.}
\end{table}

\section{Observations} \label{observations}

There are two types of H$\alpha$  observations: Fabry-Perot (F-P) spectroscopy
and narrowband imaging. Later we describe them separately, but first we discuss
issues common to both techniques.
Most of the information on the Wisconsin H-Alpha Mapper (WHAM)
in this review comes from Tufte (1997) and Haffner (1999). 
Most of the information on narrowband imaging comes from the experience
of the authors.

Both the F-P and the narrowband observations are made under clear,
dark skies, when the moon is below the horizon and the sun is more
than 18\deg below the horizon.

Both types of observations suffer from the
geocoronal emission, which is temporally and spatially variable.
The geocorona extends from a base at
an altitude of $\sim$500 km to many Earth radii.
Lyman-$\alpha$ images taken by spacecraft far above the
Earth show a smooth geocorona, decaying exponentially with geocentric
distance (Rairden {\it et al.} 1986).
We know of no such images in Lyman-$\beta$ or Balmer-$\alpha$,
but because it is so extended and because it is excited by scattered
solar Lyman-$\beta$ light, the geocorona should be rather featureless
in H$\alpha$. 
Measurements of the surface brightness $\Sigma_{geo}$
of the geocorona in H$\alpha$ over many years with instruments like
WHAM show variations from 1 to 10 R, primarily dependent on the height at which
the telescope's pointing direction exits the Earth's shadow
(Nossal {\it et al.} 1993).

For comparison the Earth-bound, dark-sky surface brightness in continuum 
${{d\Sigma_{sky}}\over{d\lambda}} = 1.3$ R \AA$^{-1}$ at
H$\alpha$  (Broadfoot \& Kendall 1968).

The two techniques use different methods to subtract geocoronal H$\alpha$
emission.
The two methods are analogous to frequency switching and
position switching.
The F-P method is analogous to frequency switching in that a
baseline is fit and subtracted as a function of frequency, so
the measured brightness at ``OFF'' frequencies affects the rest
of the spectrum.
The imaging method is like position switching (a.k.a. chopping) in
that a surface is fit and subtracted as a function of position on
the sky, so that the H$\alpha$  brightness of the sky is measured {\it relative
to} other parts of the sky.
To be exact, both methods share properties of both frequency and
position switching. In the most demanding F-P observations, the
beam is switched alternately from a target position to
an OFF position; doing so reduces the effects of time-variable
telluric emission lines. In the imaging technique, observing
alternately through an H$\alpha$  filter and a filter
near H$\alpha$ is a form of frequency switching that allows
subtraction of continuum but not line emission.

In addition to the geocorona, there are many other challenges
to H$\alpha$  observations. There are
other airglow lines, especially those from OH in the mesosphere.
The H$\alpha$  light from high velocity clouds
is within the bandpass of narrowband H$\alpha$  images but
is outside the nominal velocity coverage of the WHAM survey.
Where they have been observed, high-velocity clouds are weak, $\sim$0.1 R, or
$\sim$10\% of the H$\alpha$  surface brightness at high latitudes
(Tufte, Reynolds, \& Haffner 1998). 
There is continuum emission from the atmosphere, from zodiacal light,
from stars and galaxies, individually and collectively. Bright
stars saturate small regions of narrowband images and ruin a few
WHAM spectra by superposing their Fraunhofer H$\alpha$ absorption lines
on the spectra from the WIM. 

The Solar H$\alpha$  absorption line must also be present in the night sky 
spectrum via zodiacal light.
We are not aware of a careful analysis of this effect
on H$\alpha$  observations of the WIM.
A crude estimate is possible. The line's core is $\sim$1\AA~wide
and is nearly entirely dark ($\sim$20\% of the
continuum level). At its faintest level, the zodiacal light is
$\sim$25\% of the dark-sky continuum at H$\alpha$  (1.3 R~\AA$^{-1}$).
Thus a $\ga$0.3 R
absorption line, $\sim$50 km~s$^{-1}$ wide, should appear in all F-P spectra.
If this estimate is correct, the WHAM data should be sensitive
to this effect.
If not corrected, the effect will be to underestimate the true
brightness of the Galactic H$\alpha$  emission.

Each of the two surveys has completed its ``first year'' observations,
and these data are being calibrated and analyzed.
Both the WHAM project and the robotic imaging project at CTIO
continue to operate, reobserving some locations. Because WHAM is
remotely operated and because the CTIO project is robotically
operated, the incremental cost of reobserving is small compared to the 
original setup costs.
Once calibrated, each survey will be published independently
(anticipated dates are late-1999 to mid-2000) and then 
the region of overlap, from declination -30\deg~to
+15\deg, or 38\% of the entire sky, will be compared in detail.

\subsection{Fabry-Perot Spectroscopy} \label{fp}

Fabry-Perot H$\alpha$  spectroscopy has been dominated by
the work of Ron Reynolds, so much so that the 1-kpc thick layer of
the warm ionized medium of the Milky Way is sometimes referred to as the
``Reynolds layer.'' This has culminated in the WHAM survey
(www.astro.wisc.edu/wham/).

\begin{figure}
\plotone{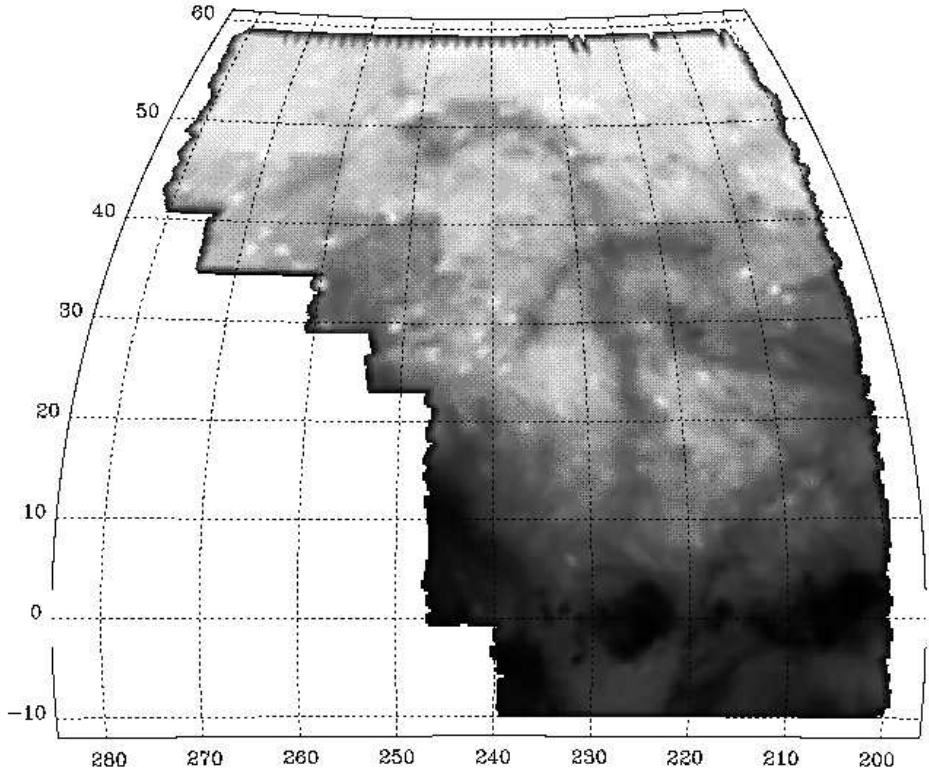}
\caption{ \label{haffner1}
An example of WHAM's survey data.
The 10\deg$\times$10\deg region centered on (l,b)=(215.8\deg,30.2\deg)
is shown at higher angular resolution in Fig. \ref{fourpanel}.
Adapted from Haffner {\it et al.} (1998), reprinted by permission.
} \end{figure}

The WHAM instrument combines a 0.6-meter
telescope and a high-resolution, 15-cm F-P spectrometer.
The spectrometer can be pressure-tuned anywhere from 4800\AA~to 7200\AA,
but here we discuss only its capabilities at H$\alpha$.
Its cryogenically-cooled CCD has high quantum efficiency (78\% at H$\alpha$) and
low noise (3 e- rms). 
It is located at Kitt Peak, AZ, and operated remotely from Wisconsin.
Fig. \ref{haffner1} shows an
example of WHAM's view of the WIM.

\begin{figure}
\plotfiddle{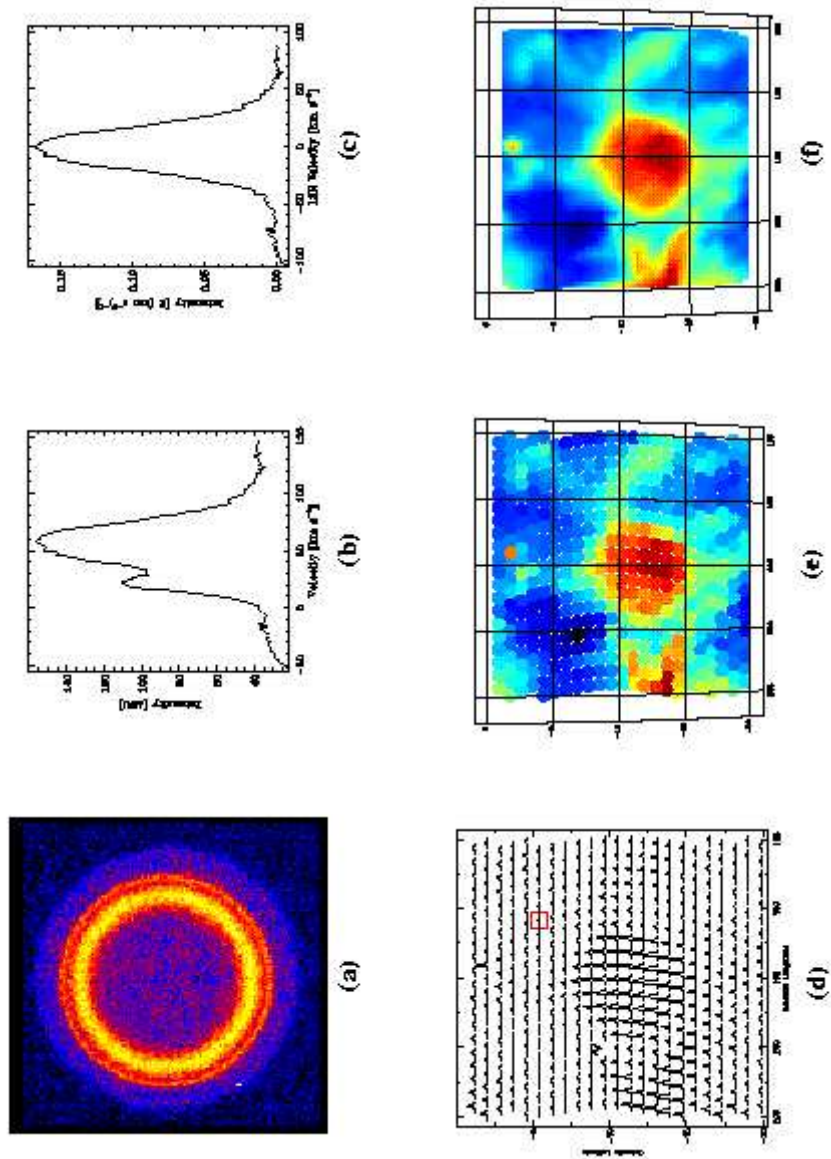}{14truecm}{0}{64}{64}{-195}{-30}
\caption{ \label{haffner2}
Steps in processing WHAM data.
(a) The CCD data from one beam on the sky. Each ring corresponds to
line emission at a different velocity. The radial velocity decreases
from center to edge. 
(b) The spectrum is formed by azimuthally averaging the CCD data from (a).
Velocity coverage is $\sim 200$ km~s$^{-1}$, with resolution 12 km~s$^{-1}$.
(c) The calibrated spectrum includes flat fielding, baseline fitting,
intensity and velocity calibration, and subtraction of the geocoronal line.
(d) A grid of calibrated spectra.
A small square highlights the example spectrum in the previous panels.
(e) A grid of nested beams show the integrated light within the range
V$_{LSR} = -6$ to $+6$ km~s$^{-1}$. 
(f) The final, rectilinear image results from oversampling and smoothing
the data from panel (e).
Reproduced in B\&W, intensity does not map monotonically to grey scale.
From Haffner (1998), reprinted by permission.
} \end{figure}

The WHAM survey is a velocity-resolved survey
of H$\alpha$ emission over 3/4 of the entire sky (declination $>
-30$\deg). The survey consists of more than 37,000 individual
spectra, each with velocity resolution of 12 km~s$^{-1}$ over a 200 km~s$^{-1}$
(4.4\AA) spectral window. 
Each spectrum is observed for 30 seconds, yielding a 3$\sigma$
sensitivity of 0.15 R. 
The 1\deg~diameter beam determines the
angular resolution of the survey. The beams are spaced by 
$\Delta$b = 0.85 deg and $\Delta$l = 0.98 deg / cos(b). The beams
are nested in a close-packed arrangement like racked billiard balls.
That is, adjacent rows along constant latitude are offset by one half
a beam spacing in longitude. The spherical geometry of the sky
necessitates some ``defects'' in this close-packed lattice 
(see Fig \ref{haffner2}(e)).

Because the geocorona is at rest with respect to the Earth,
geocoronal H$\alpha$  peaks near zero velocity with respect
to the ground-based observer. By using the Earth's 30 km~s$^{-1}$
velocity around the Sun and allowing for the Sun's velocity with
respect to the local standard of rest (LSR), one can maximize
the separation in observed velocity between Galactic and geocoronal H$\alpha$.
The observing schedule for the WHAM survey was designed such that
the geocoronal H$\alpha$  line was centered at $|{\rm V_{LSR}}| \ga 20$ km~s$^{-1}$,
except at the north ecliptic pole where the maximum possible separation is 15
km~s$^{-1}$.


Marcelin {\it et al.} (1998) used F-P spectroscopy to 
show that in the SCP region, free-free emission is not
a significant contributor to the observed microwave anisotropy.
The SCP results are in concordance with previously reported observations
180\deg away near the NCP made with imaging techniques 
(Gaustad, McCullough, \& Van Buren 1996; Simonetti, Dennison, \& Topasna 1996).
Figure 4 shows a spectrum from Marcelin {\it et al.} (1998).
Note the careful component-- and baseline-fitting
required to extract the Galactic H$\alpha$  emission. The Galactic H$\alpha$~
in Figure 4 is 1.0 R, dwarfed by the 10 R line of geocoronal H$\alpha$.
For comparison, the WHAM spectrum in Figure \ref{haffner2}(b)
shows a 7 R Galactic H$\alpha$  line with a much weaker, $\sim$2 R geocoronal
line on its left wing.
This illustrates the variability of the geocorona (see also
Figure 1 of Bartlett \& Amram (1998)).

The F-P technique can also be used as a narrowband imager.
WHAM can be reconfigured
to provide 2\arcmin~angular resolution images of faint emission
line sources within WHAM's 1\deg~diameter circular beam. When used
at the highest spectral resolution (12 km~s$^{-1}$), an emission line source
of 1 R and 2\arcmin~angular extent can be detected with a signal
to noise ratio of five in a 1000 second integration (Reynolds 1998).
Since the original WHAM survey required one year of observing
at 30 seconds per 1\deg~beam, an all-sky survey with WHAM
at 2\arcmin~resolution would require considerable patience, although
smaller regions would be feasible.

To achieve all-sky H$\alpha$  observations with arcminute resolution
in a reasonable amount of time, another solution is the narrowband
imaging method, discussed in the next section.

\begin{figure}[p]
\plotone{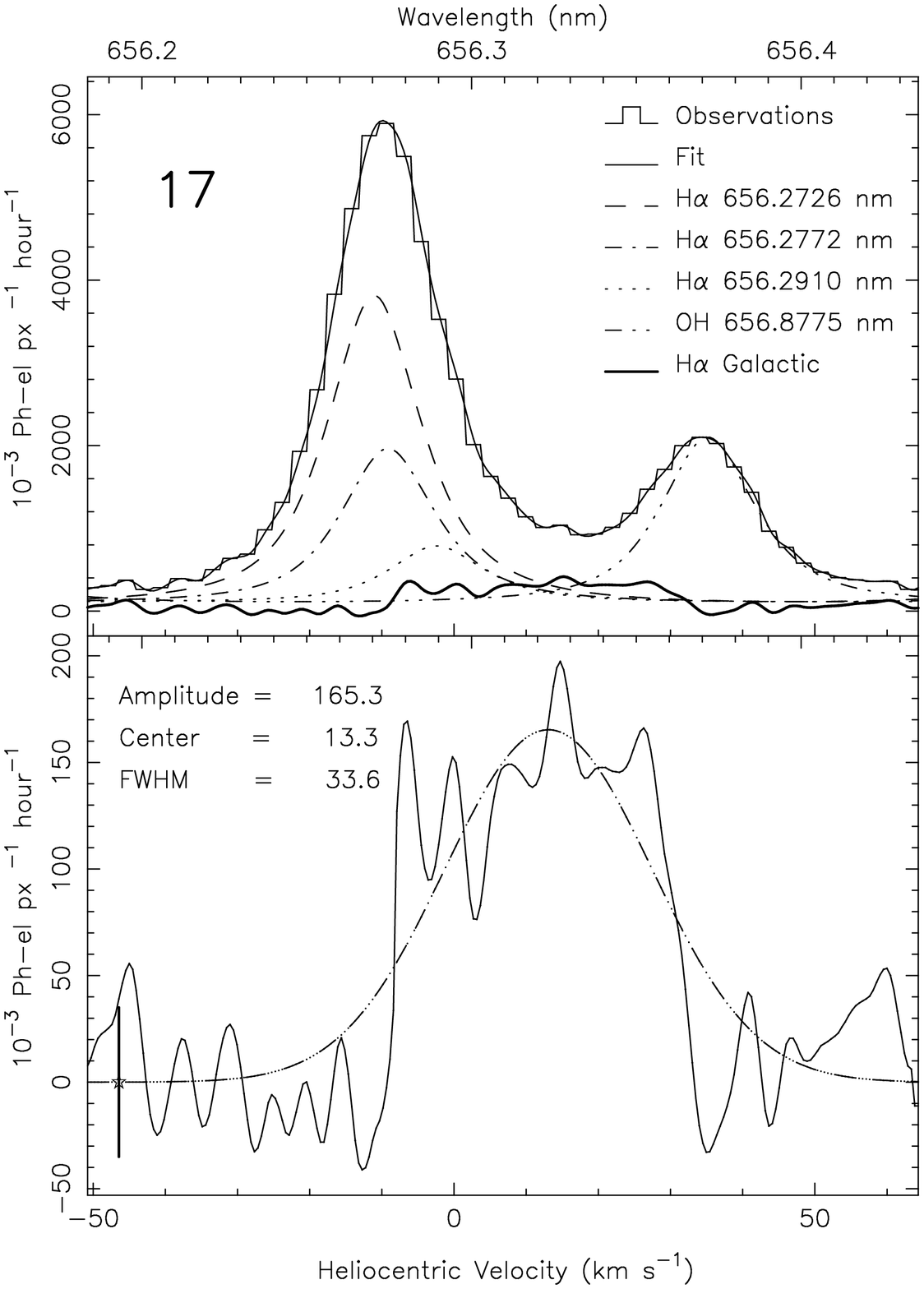}
\caption{ \label{marcelin}
Top: Careful component-- and baseline-fitting is
required to extract the Galactic H$\alpha$  emission from Fabry-Perot
spectra. The Galactic H$\alpha$  spectrum is the thick solid line.
Bottom: The Galactic H$\alpha$  spectrum is plotted again with a 20X
expanded scale. The Galactic H$\alpha$  emission is 1.0 R, typical for high
latitude regions. A Gaussian fit is plotted too.
Reprinted by permission from Marcelin {\it et al.} (1998).
} \end{figure}

\subsection{Narrowband Imaging} \label{imaging}

To make wide-field, narrowband images,
an interference filter can be placed in front of
a camera lens attached to a CCD. The interference filter
typically is 10 to 50\AA~wide (FWHM). A narrow bandpass $\Delta\lambda$
is desirable to suppress continuum light from the night sky, 
${{d\Sigma_{sky}}\over{d\lambda}}$,
because the signal-to-noise ratio SNR of the H$\alpha$  light
detected in time $\tau$ over a solid angle $\Omega$ 
is photon-limited if detector noise can be neglected:
\begin{equation}
SNR \la \tau~\Omega~\Sigma_\alpha / \sqrt{\tau~\Omega~(\Delta\lambda~{{d\Sigma_{sky}}\over{d\lambda}}~+ \Sigma_\alpha)}.
\end{equation}
However, the narrower
the bandpass, the narrower is the usable field of view,
because the central wavelength $\lambda$ of an interference filter shifts
blueward by ${\delta}\lambda$ as the angle of incidence $\theta$ increases,
\begin{equation}
{\delta}\lambda \approx - \lambda~{{sin^2(\theta)}\over{2~n^2}},
\end{equation}
where $n$ is the index of refraction of the filter ($n\approx 2$).
Equation 6 implies that 
the usable solid angle $\theta^2$ is approximately proportional to filter
width $\Delta\lambda$.
Equation 5 implies that to detect H$\alpha$  signals that are much fainter
than the continuum, the time required is proportional to
the filter width: $\tau \propto \Delta\lambda$.
Thus, a narrowband imaging survey can be made as quickly with a wide
filter and wide field of view as a narrow filter and narrow field of
view. 
Wide filters require extremely precise gain
calibration (i.e. flat-fielding) to see the small fraction
of the light that is H$\alpha$. 
On the other hand, narrow fields of view result in more images to mosaic.
This analysis breaks down at very wide filters ($\Delta\lambda
\ga 50$~\AA) because telluric emission lines (primarily those of OH
at 6554 \AA~and 6577 \AA) increase the brightness of the sky.
Similarly, the analysis
breaks down for very narrow filters ($\Delta\lambda \la 5$~\AA)
because the geocoronal H$\alpha$  emission begins to
exceed the sky continuum
($\Sigma_{\alpha, geo} \ga ~\Delta\lambda{{d\Sigma_{sky}}\over{d\lambda}}$).

\begin{figure}[p]
\plotfiddle{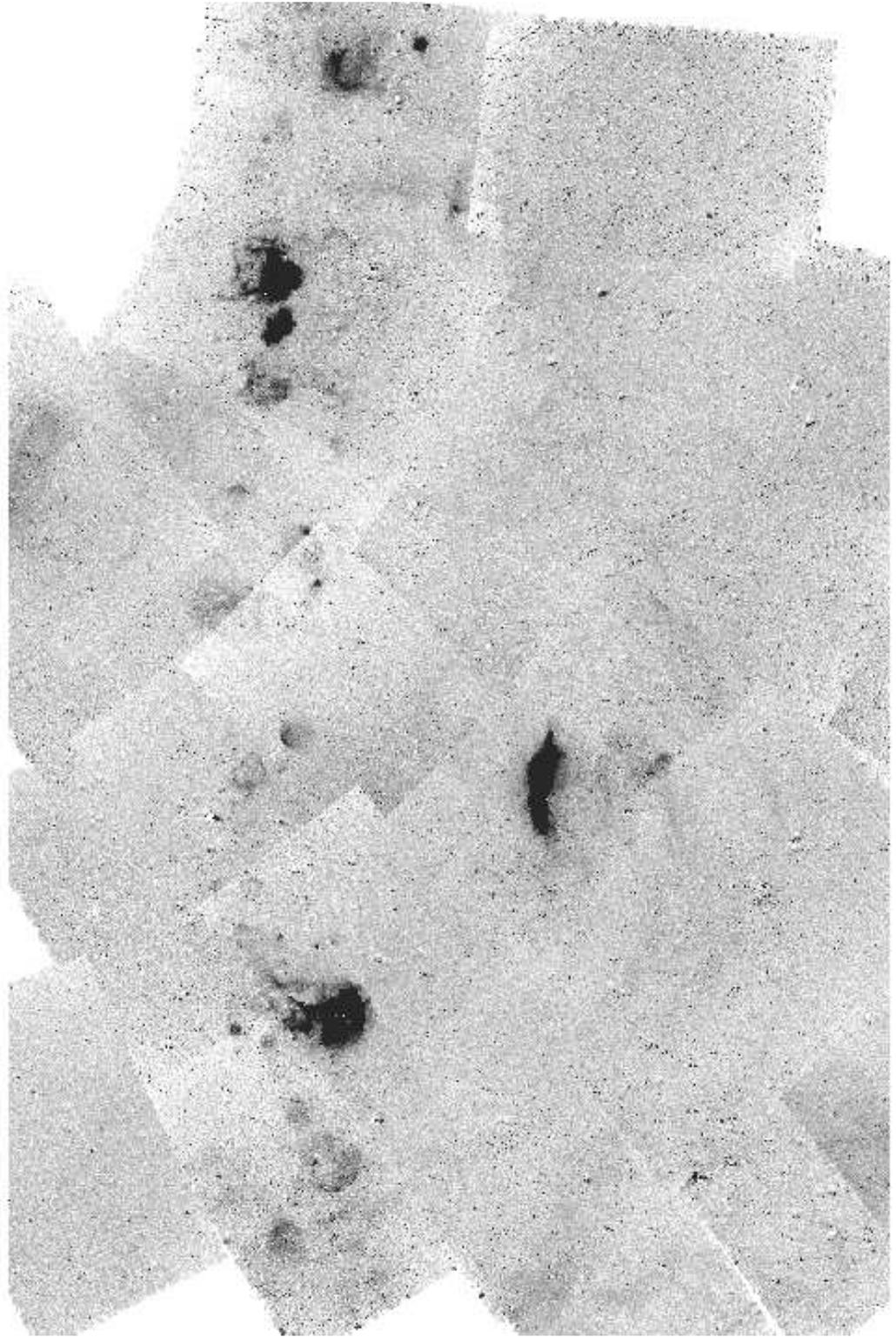}{18truecm}{0}{80}{80}{-250}{-20}
\caption{ \label{anticenter}
Imaged in H$\alpha$, the Galactic anticenter region shows classic Stromgren
spheres in the Galactic plane, and filaments of ionized gas at 
moderate latitudes. 
The region displayed extends from  l = 120\deg~to 190\deg,
b = -30\deg~to +15\deg and
the limiting sensitivity is 4 R at 2\arcmin~resolution.
Adapted from McCullough, Reach \& Treffers (1990).
} \end{figure}

The narrowband imaging technique was applied
to a large field of view by McCullough, Reach, \& Treffers (1990) to
create a 70\deg$\times$45\deg H$\alpha$  mosaic near the Galactic anticenter
(Fig. \ref{anticenter}). They used a
$320\times512$ pixel RCA CCD, a 50-mm lens, and a 44\AA~wide
H$\alpha$  filter at a light-polluted site.
Their wide-field H$\alpha$  mosaic
revealed H~II regions in the Galactic plane and 
filamentary structures at moderate Galactic latitudes.
The mosaic also shows the edges of the individual
images; they are prominent because the median value of
each image was forced to zero by subtracting a constant, rather
than a more sophisticated method such as globally fitting and
individually subtracting a set of polynomials matched to the
background of each image.

Gaustad, Rosing, McCullough, \& Van Buren (1998) have
installed a robotic camera at Cerro Tololo Inter-american Observatory
in Chile (www.astronomy.
swarthmore.edu).
The robotic mount and dome point and protect the camera
with almost zero assistance. In nominal operation, the robot requires
human interaction for three operations: 1) if its weather monitor
indicates that conditions are fine to open the dome, the robot confirms
that with the 4-m telescope operator by email, 2) the 4-m operator also may
tell the robot to close its dome at any time if conditions warrant, and
3) each week a human inserts a blank tape and mails a full tape to the USA for
analysis.

The thermoelectrically cooled 1024$\times$1024 CCD
has 12~\micron~pixels. Coupled to a Canon lens of 50-mm focal length,
the CCD yields a field of view of 13\deg$\times$13\deg
and a scale of 0.8\arcmin~per pixel. (The 30-mm diameter aperture is
similar to the apertures of Galileo's first telescopes!)
Images are taken through
a 30\AA~wide H$\alpha$  filter and a dual-bandpass filter that excludes
H$\alpha$  but transmits two 60\AA~bands of continuum light, one on each
side of H$\alpha$. The filters are enclosed in an insulated, thermostated wheel
to prevent drifts in the transmission curves. 
Because twilight sky is inadequately flat over wide fields of
view (Chromey \& Hasselbacher 1996), flat-fielding is accomplished
with a custom light box with both incandescent and H$\alpha$  lamps.

The robotic imaging survey consists of 269 fields covering the sky 
from declination -90\deg~to +15\deg, with the same centers as those in the
IRAS Sky Survey Atlas. Each field is observed six times in continuum
(5 minutes per exposure) and five times in H$\alpha$  (20 minutes
per exposure). The H$\alpha$  images are interleaved between the
continuum images to minimize differences in the continuum level between
the two as the zenith angle changes.
Multiple exposures also permit rejection of cosmic rays, airplanes,
and satellites.
Figure 6 illustrates the image processing.
Sensitivity is maximized at angular scales from 0.1\deg~to 2\deg, with
much smaller features being limited by shot-noise and residuals of
star-subtraction, and much larger features being difficult to
distinguish from variations in the foreground sky and flat-fielding
errors. Structures of angular extent 0.1\deg~to 2\deg 
and surface brightnesses of $\ga$1 R are detected reliably
in the final images. 

Because the 13\deg$\times$13\deg~images are spaced by 10\deg, every
spot within 3\deg~of the edge of any image appears on an adjacent
image also (the corners appear 4 times). This redundancy is helpful in
mosaicing images and in recognizing genuine Galactic H$\alpha$  emission.
Some fields were missed in the first year of observing. Those fields are
the highest priority of the second year, with the remaining time allocated
to observing a second grid of positions offset from the first grid
by 5\deg. As of February 1999, 234 (87\%) of the fields had been observed,
with most of the missing fields at declination $\delta \geq 0$\deg.

Another imaging survey has been conducted by 
Dennison, Simonetti, \& Topasna (1998). Their cryogenically-cooled CCD camera
has a 58-mm lens, 27~\micron~pixels, 
a scale of 1.6\arcmin/pixel, and covers a
13.6\deg$\times$13.6\deg field of view.
They have observed much of the Milky Way ($|b| \la 30$\deg)
from southwestern Virginia.

Galactic plane or low-latitude observations probably are not relevant to
the CMB-foreground problem only because all foregrounds are too
strong at low latitudes to permit templates to be subtracted accurately.
The H$\alpha$  survey of Buxton, Bessell, \& Watson (1998)
and the Br$-\gamma$ (2.1655~\micron) observations of
Kutyrev {\it et al.} (1997) are in that category.
However, Br$-\gamma$ is a good alternative to H$\alpha$~
wherever dust extinction is significant, i.e. primarily at low latitudes.

\begin{figure}[p]
\plottwo{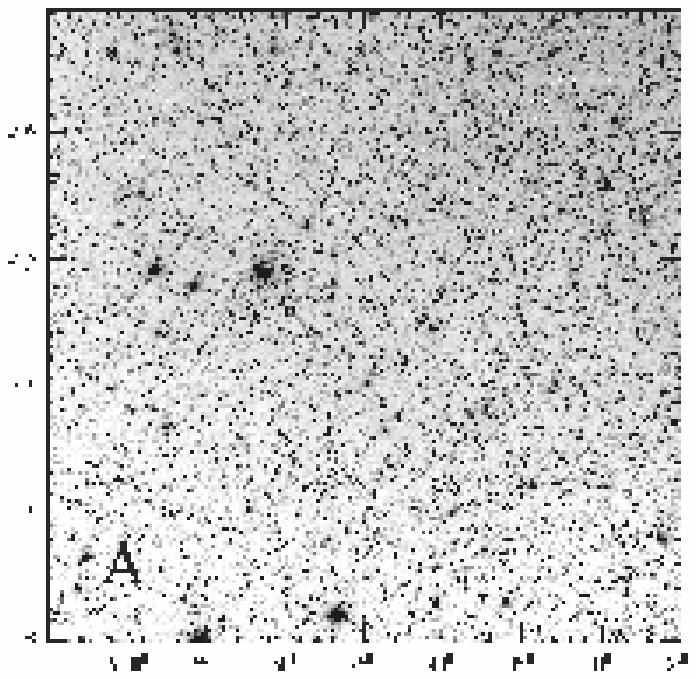}{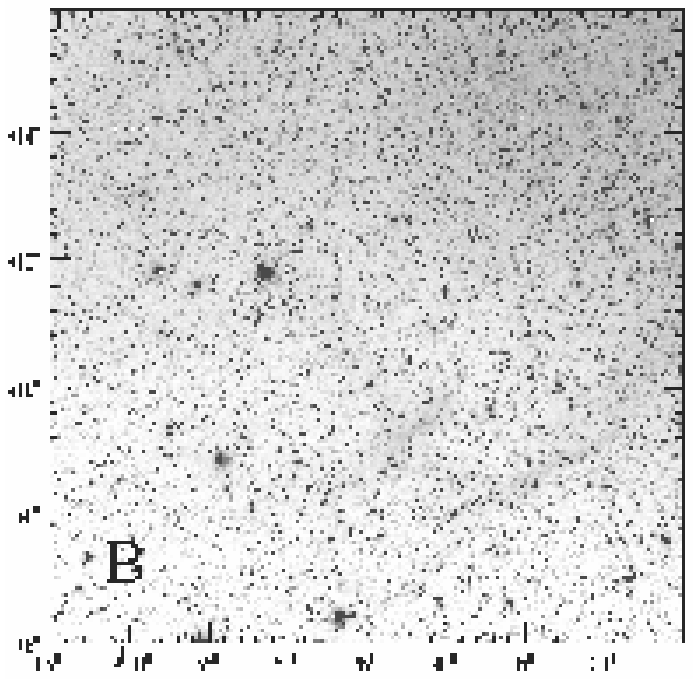}
\vskip 0.01truecm
\plottwo{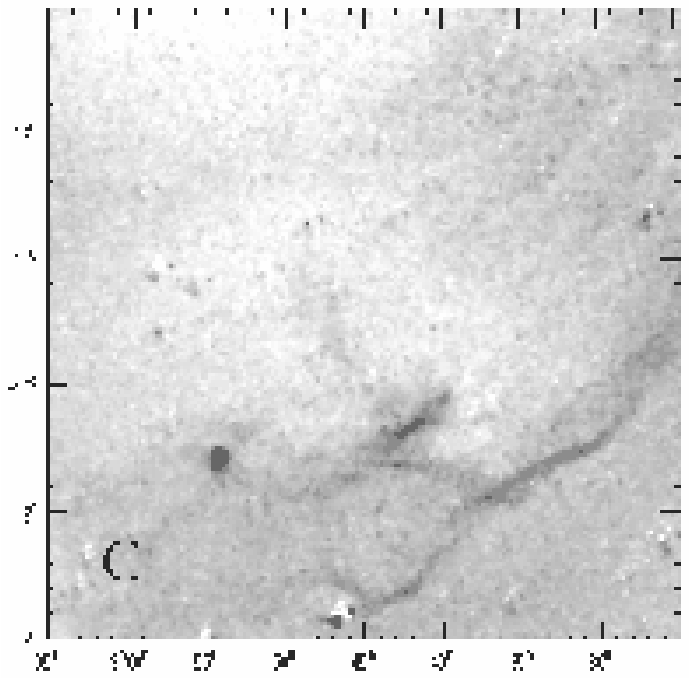}{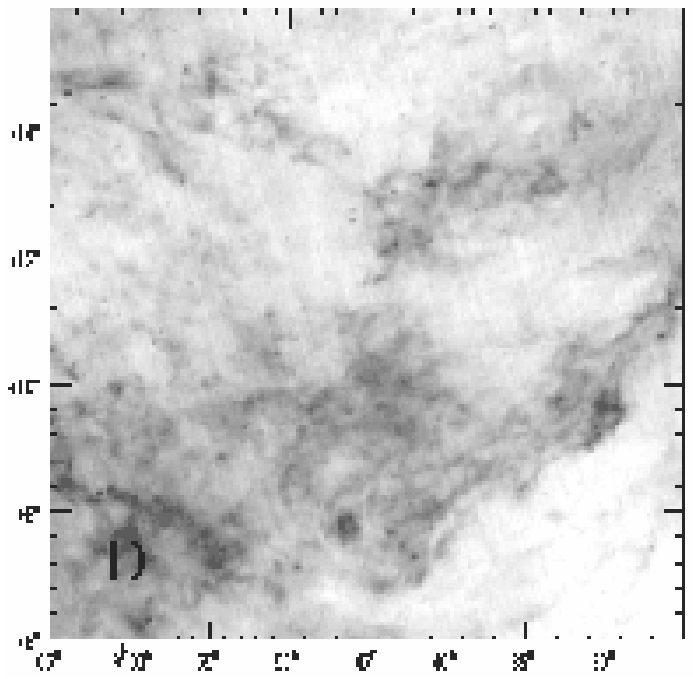}
\caption{ \label{fourpanel}
The data processing steps for the imaging technique are outlined here
with data from CTIO.
A) A flat-fielded image made with a continuum filter.
B) Same as A but with an H$\alpha$  filter.
C) The difference image shows only H$\alpha$  emission and residuals
around a few bright stars. This difference image results from
1) multiplying image A by a constant to match the
stellar photometry of the two images A \& B, 2) registering and
subtracting the scaled image A from image B, 3) median filtering
the result with a 7x7 pixel (5\arcmin$\times$5\arcmin) kernel to suppress
residuals around stars,
and finally fitting and subtracting an arbitrary first-order surface
(a plane) to remove geocoronal emission.
The bright source at 8\fh55\fm +9\deg is the planetary nebula A 31.
D) The 100~\micron IRAS ISSA image of the same region. The tendency for the
H$\alpha$  emission to outline the dust emission is discussed in
Section \ref{correlations}
The Galactic plane is toward the lower right (southwest).
} \end{figure}

\section{Correlations} \label{correlations}

Dust and gas are well mixed in the ISM, so we might expect to see
a positive correlation between dust emission and free-free emission.
If instead the dust and gas in the ISM were like immiscible fluids such
as oil and water, we might expect to see an anti-correlation, namely
that where there is much dust (oil), there is little gas (water).

Some models of the ISM predict that neutral atomic
hydrogen gas and ionized hydrogen gas are not well mixed but may be
spatially and physically associated. For example,
the model of McKee \& Ostriker (1977) has H I clouds with H II
skins surrounding them. As discussed below, H II skins around clouds
are observed.

Reynolds {\it et al.} (1995) examined the relationship between H I and H II
in velocity-resolved maps of H$\alpha$  and 21-cm emissions. They observed
H$\alpha$~-emitting H I clouds and for those clouds measured
the ratio I$_{\alpha}$/N$_{H I} \approx 20$ R/10$^{21}$ cm$^{-2}$.
They also noted that the ratio decreased systematically from 
$\sim 15$ R/10$^{21}$ cm$^{-2}$ to 
$\sim 4$ R/10$^{21}$ cm$^{-2}$ as the VLSR increased from -50 km~s$^{-1}$ to
0 km~s$^{-1}$, corresponding to a decrease in distance from
the Galactic plane, $|z|$, from 1 kpc to 0 kpc. Converting from
I$_{\alpha}$/N$_{H I}$ to I$_{\alpha}$/I$_{100}$ with the factor
${\rm I_{100}/N_H = 8.5~MJy~sr^{-1} / 10^{21} cm^{-2}}$
typical for high latitude regions (Boulanger \& Perault 1988),
we find that Reynolds {\it et al.}'s results predict a range in slopes for
the correlated components of H$\alpha$  and 100~\micron~emissions
from 0.5 to 2.0 R / (MJy~sr$^{-1}$). 
Kogut (1997) directly compared
Reynolds {\it et al.}'s H$\alpha$  data with infrared data and derived
an average ratio of
I$_{\alpha}/{\rm I_{100}} = 0.85\pm0.44$ R / (MJy~sr$^{-1}$), which
is within the range previously derived using the Boulanger \& Perault
factor.

Reynolds {\it et al.} found that the H$\alpha$~-emitting H I clouds account
for only 30\% of the H$\alpha$  emission and only 10\% of the
21-cm emission. The amount of correlated emission is
much less than the amount of uncorrelated emission.
That is, the degree of correlation between
H$\alpha$  and 21-cm emissions is not strong; it is weak, at
angular scales less than a few degrees.

\begin{figure}[p]
\plotone{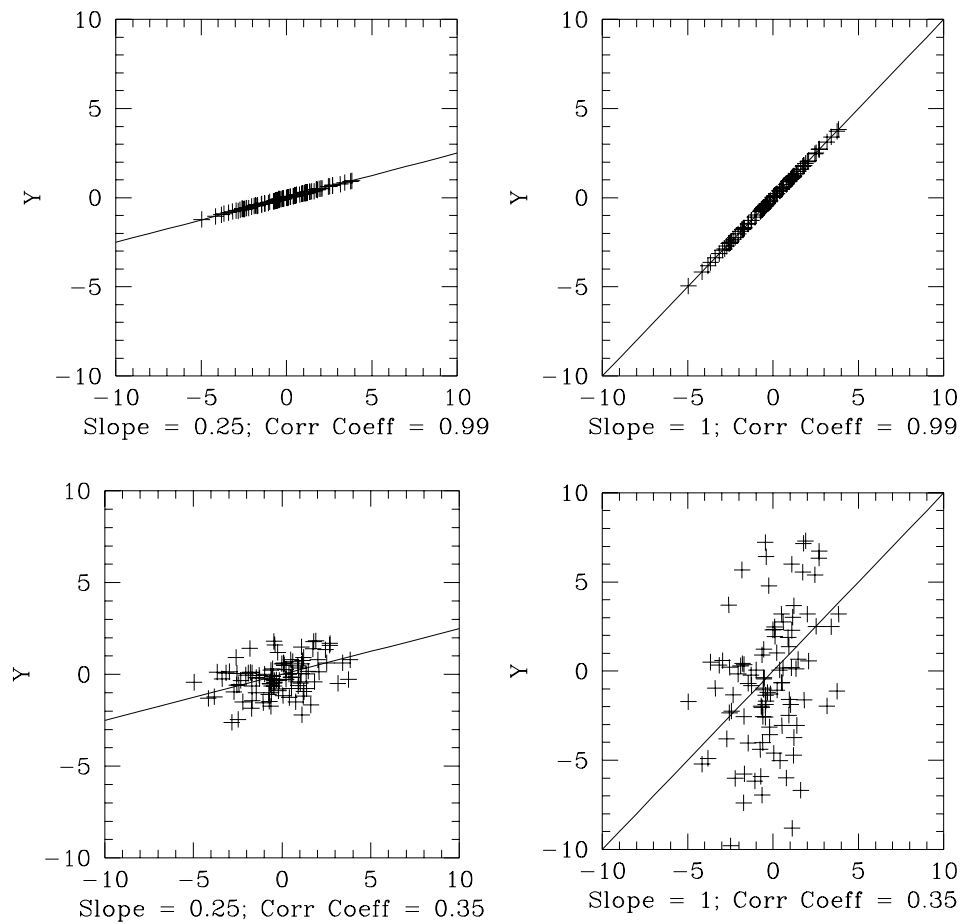}
\caption{ \label{fig-correlations}
The correlation between H$\alpha$  (or free-free) and
dust emission is discussed in Section \ref{correlations}
In this figure, pairs of correlated random numbers illustrate the meaning of
the degree (or strength) of a correlation and its slope.
The label lists the
correlation's slope and its Pearson correlation coefficient.
The latter quantifies the degree (or strength) of the correlation.
The correlation is good (or strong) in the top panels and poor (weak)
in the lower panels.
The slope is four times larger in the right panels than the left panels.
(To some people the fit in the lower left panel looks more accurate than the fit
in the lower right panel, but that is an optical illusion: the data are 
identical except for a scale factor in Y.)
} \end{figure}

Figure \ref{fig-correlations} illustrates the meaning of the degree
(or strength) of a correlation as compared with the slope of that correlation.

\begin{table}
\caption{H$\alpha$ and Free-free Correlations with Dust\tablenotemark{a}} \label{tbl-2}
\begin{center}
\begin{tabular}{ccll}
${\rm I_{\alpha}/I_{100} [R / (MJy~sr^{-1})]}$ & Type & Source & Notes\\
\tableline
$0.34\pm0.33$ 		&H$\alpha$~		&	Kogut (1997)&b\\
$0.79^{+0.44}_{-0.15}$ 	&H$\alpha$~		&	McCullough (1997)&b\\
$0.85\pm0.44$ 		&H$\alpha$~		&	Kogut (1997)&b\\
$0.5-2.0$ 		&H$\alpha$~		&	Reynolds {\it et al.} (1995)&c\\
$2.0\pm0.5$ 		&microwave	&	Kogut {\it et al.} (1996)&b\\
$3.6\pm2$ 		&microwave	&	de Oliveria-Costa {\it et al.} (1997)	&b\\
$6$ 			&microwave	&	Leitch {\it et al.} (1997)&b,c\\
$37\pm19$ 		&microwave	&	Lim {\it et al.} (1996)&d\\
\end{tabular}
\end{center}
\tablenotetext{a}{Microwave results converted to H$\alpha$  equivalents assuming $T_e = 10^4$K.}
\tablenotetext{b}{As compiled by Smoot (1998).}
\tablenotetext{c}{See text for discussion.}
\tablenotetext{d}{As reported by Kogut (1997).}
\end{table}

Smoot (1998) has collected estimates of the slope of the
correlation of H$\alpha$  and dust, and of free-free and dust;
those are reproduced in Table 2.
The aforementioned result,
I$_{\alpha}/{\rm I_{100}} = 0.5$ to 2.0 R / (MJy~sr$^{-1}$),
and a result of Lim {\it et al.} (1996) are included also.

From data like those in Table 2,
it has been noted that the H$\alpha$  observations imply 
less radio emission per unit dust emission than is observed
in microwaves (Kogut 1997; Smoot 1998).
If true, this might be evidence of an additional mechanism for
microwave emission emanating from dust, perhaps of the sort
described by Lazarian (in this volume) and plotted in Figure 1.

We can test the hypothesis that a single
value for the ratio I$_{\alpha}/{\rm I_{100}}$ is consistent with
the data in Table 2. In doing so, we assume the measurements
are not biased and that the error distributions are normal (a.k.a. Gaussian).
As a practical matter, we do not include results that lack formal error
estimates.\footnote{ For this reason, the Leitch {\it et al.} (1997) and the
Reynolds {\it et al.} (1995) results are not included. Furthermore, the region
near the NCP studied by Leitch {\it et al.} is atypically cool and therefore
deficient in 100~\micron~flux per unit dust column density by a factor
of $\sim$3 (Finkbeiner 1999).}
The maximum likelihood value is 0.86 ($\pm$0.2) R / (MJy~sr$^{-1}$).
No individual result differs from the maximum likelihood value by more than
2.3 times the respective standard deviation. For the 6 measurements and
1 degree of freedom, the minimum $\chi^2 = 13.2$, which
occurs 2\% of the time by chance alone.
The Lim et al (1996) result may be affected by anisotropy in the CMB
(Kogut 1997), and if we disregard it, then for the 5 remaining measurements,
the minimum $\chi^2 = 9.6$, which occurs 5\% of the time by chance alone.
Therefore, these data do not refute the hypothesis convincingly, especially
if one considers the additional possibilities of bias in the means
and non-Gaussian errors.

Kogut {\it et al.} (1996) report a good correlation
between free-free emission and dust on $\ga 7$\deg~angular scales observed
with COBE's DMR. McCullough (1997) reports a weak correlation
between H$\alpha$  emission and dust on $\la 1$\deg~angular scales.

That the correlation weakens with decreasing angular scale can be
understood with a simple geometric model in which neutral dusty clouds
are externally ionized from a source emanating from the Galactic plane.
A good example is shown in Figure 6C and D. The
dust cloud has a ``silver lining'' of H$\alpha$  emission on the side of
cloud nearest the Galactic plane. The same phenomenon is evident in
another cloud at higher latitude (McCullough 1997). 

The external ionization creates a skin of ionized gas preferentially
on the side of the cloud nearest the Galactic plane. The narrowness of
the H$\alpha$  lining indicates that the source of ionization doesn't
penetrate far into the cloud. 
(The 100~\micron~surface brightness
implies A$_V \leq 0.2$ mag, so H$\alpha$  light
can pass through the cloud with minor attenuation.)
If the surface of a cloud is uniformly ionized, for example by
external photoionization, the ``silver lining'' effect will occur wherever the
cloud's surface is tangent to the line of sight, due to the
longer path length.
This model predicts that the degree of correlation between dust and
ionized gas (observed by H$\alpha$  or free-free emission)
will be A) weak on angular scales smaller than the
size of a typical cloud but B) stronger on much larger angular scales.

Haffner {\it et al.} (1998) report WHAM data showing H$\alpha$  emission with
little if any association with other emissions such as
X-ray, infrared, or 21-cm. On the whole, narrowband images support this view,
namely that in many cases, there is no obvious relationship between
ionized gas and other constituents. However, in some case such as
those discussed in the previous paragraphs, the detail revealed in
the narrowband images clearly show a physical association.
In other cases, the velocity revealed by spectroscopy is critical
to our understanding. 

\section{Conclusions} \label{conclusions}

Halpha observations completed in the past year will
provide a template of the free-free emission from the Milky Way.
The H$\alpha$  sky has been observed with a high-sensitivity Fabry-Perot
spectrometer from the northern hemisphere, and with a narrowband imager
from the southern hemisphere.
Approximately 38\% of the H$\alpha$  sky has been observed with both
instruments.
Both surveys are sufficiently sensitive to detect
warm ionized hydrogen gas at all Galactic latitudes.
A simple geometric model is proposed to explain observed correlations between
ionized gas and dust.

Astronomers interested in the CMB will create
maps that are free of free-free for free (i.e. without paying
a noise penalty inherent in multi-frequency estimates of free-free emission).
Alternatively, it may be possible
to quantify the location and the brightness of another foreground
such as that from rotating dust grains by comparing
the H$\alpha$  template with the templates of synchrotron, 
free-free (directly-observed in microwaves), and thermal dust.

%
%

\acknowledgments

We have benefited from the help of the WHAM team, in particular Matt Haffner,
Ron Reynolds, and Steve Tufte.
Students that have assisted us are 
Bender,
Chen,
Farney,
Hall,
Hentges,
Khosrowshahi,
Logan,
Oh,
Pulokas,
Raschke,
Schneider,
Seaton,
and
Tajima.
We also appreciate the assistance of the staff of CTIO in installing
and maintaining our robot.
The authors' research has been supported by grants from the National
Science Foundation, Las Cumbres Observatory, Dudley Observatory,
the Fund for Astrophysical Research, the Research Corporation, NASA,
Swarthmore College, and the University of Illinois at Urbana-Champaign.
This work was performed in part at the Jet Propulsion Laboratory of
the California Institute of Technology, which is operated under contract
with the National Aeronautics and Space Administration.

%
%

\end{document}